\documentclass[10pt,conference]{IEEEtran}
\usepackage{amsfonts,amsmath,amssymb}
\usepackage{graphicx,color,epsfig,rotating}
\usepackage{latexsym}
\usepackage{subfigure}
\usepackage{cite}

\long\def\comment#1{}

\newfont{\bbb}{msbm10 scaled 700}

\newfont{\bb}{msbm10 scaled 1100}
\newcommand{\CC}{\mbox{\bb C}}

\newcommand{\RR}{\mbox{\bb R}}

\newcommand{\EE}{\mbox{\bb E}}

\newcommand{\bv}{{\bf b}}
\newcommand{\cv}{{\bf c}}
\newcommand{\dv}{{\bf d}}

\newcommand{\gv}{{\bf g}}
\newcommand{\hv}{{\bf h}}

\newcommand{\pv}{{\bf p}}
\newcommand{\qv}{{\bf q}}
\newcommand{\rv}{{\bf r}}
\newcommand{\sv}{{\bf s}}
\newcommand{\tv}{{\bf t}}

\newcommand{\vv}{{\bf v}}
\newcommand{\xv}{{\bf x}}
\newcommand{\yv}{{\bf y}}
\newcommand{\zv}{{\bf z}}
\newcommand{\zerov}{{\bf 0}}
\newcommand{\onev}{{\bf 1}}

\newcommand{\Am}{{\bf A}}
\newcommand{\Bm}{{\bf B}}
\newcommand{\Cm}{{\bf C}}

\newcommand{\Gm}{{\bf G}}
\newcommand{\Hm}{{\bf H}}
\newcommand{\Id}{{\bf I}}

\newcommand{\Rm}{{\bf R}}

\newcommand{\Tm}{{\bf T}}
\newcommand{\Um}{{\bf U}}

\newcommand{\Cc}{{\cal C}}

\newcommand{\Lc}{{\cal L}}

\newcommand{\Nc}{{\cal N}}

\newcommand{\Sc}{{\cal S}}

\newcommand{\gammav}{\hbox{\boldmath$\gamma$}}

\newcommand{\lambdav}{\hbox{\boldmath$\lambda$}}

\newcommand{\Sigmam}{\hbox{\boldmath$\Sigma$}}
\newcommand{\Phim}{\hbox{\boldmath$\Phi$}}

\newcommand{\trace}{{\hbox{tr}}}

\newcommand{\eqdef}{\stackrel{\Delta}{=}}

\newcommand{\herm}{{\sf H}}

\newcommand{\transp}{{\sf T}}

\newcommand{\xsf}{{\bold{\sf x}}}

\begin{document}

\title{MIMO Broadcast Channel Optimization \\ under General Linear Constraints }
\author{\IEEEauthorblockN{Hoon Huh\IEEEauthorrefmark{1}, Haralabos Papadopoulos\IEEEauthorrefmark{2}, Giuseppe Caire\IEEEauthorrefmark{1}}
\IEEEauthorblockA{\IEEEauthorrefmark{1}University of Southern California, Los Angeles, CA, Email: hhuh, caire@usc.edu}
\IEEEauthorblockA{\IEEEauthorrefmark{2}DoCoMo Laboratories USA, Inc., Palo Alto CA, Email: hpapadopoulos@docomolabs-usa.com}
}
\maketitle

\begin{abstract}
The optimization of the transmit parameters (power allocation and steering vectors) for the MIMO BC under general linear constraints is treated under the optimal DPC coding strategy and the simple suboptimal linear zero-forcing beamforming strategy. In the case of DPC, we show that ``SINR duality'' and ``min-max duality''
yield the same dual MAC problem, and compare two alternatives for its efficient solution.
In the case of zero-forcing beamforming, we provide a new efficient algorithm based on the direct optimization
of a generalized inverse matrix. In both cases, the algorithms presented here address the problems in the most general form and can be applied to special cases previously considered, such as per-antenna and per-group of antennas
power constraints, ``forbidden interference direction'' constraints, or any combination thereof.
\end{abstract}

\section{Model and background} \label{sec:model}

One channel use of the MIMO BC with an $M$-antenna transmitter and $K$ single-antenna receivers is defined by
\begin{equation} \label{general-down}
y_k = \hv_k^\herm \xv + z_k, \;\;\; k = 1,\ldots, K
\end{equation}
where $\hv_k, \xv \in \CC^M$ are the channel vector of user $k$ and the transmitted signal vector, respectively,
and $z_k \sim \Cc\Nc(0,1)$ is AWGN.
The relevance of the above model for the downlink of a wireless system has been widely discussed.
Also, the impact of non-ideal channel state information and practical techniques for channel estimation and channel state feedback
are well-understood (see for example \cite{Caire-Jindal-Kobayashi-ISIT07, Caire-Jindal-Kobayashi-ISIT08} and references therein).
Here, we assume fixed channel vectors perfectly known to all
terminals
and focus on the optimization of the transmitter parameters.

Let $\Sc$ denote a {\em compact} set of $M \times M$ covariance matrices.
The capacity region of the MIMO BC (\ref{general-down}) subject to the input constraint $\EE[ \xv \xv^\herm] \eqdef \Sigmam_x \in \Sc$ is given by the set of
rate points $\Rm \in \RR_+^K$ \cite{Weingarten-Steinberg-Shamai-TIT04}
\begin{eqnarray} \label{cap-region}
 \Cc(\Sc; \hv_1,\ldots, \hv_K) = \mbox{coh} \; \bigcup_{\sum_{k=1}^K \vv_k \vv_k^\herm q_k  \in \Sc} \;\; \bigcup_{\pi} & & \nonumber \\
\left \{ R_{\pi_k} \leq \log \left (1 + \frac{|\hv_{\pi_k}^\herm \vv_{\pi_k}|^2 q_{\pi_k}}{1 +
\sum_{j=k+1}^K |\hv_{\pi_k}^\herm \vv_{\pi_j}|^2 q_{\pi_j}} \right ), \;\; \forall \; k \right \} & & \nonumber \\
& &
\end{eqnarray}
and it is achieved by Dirty-Paper Coding (DPC) where the permutation $\pi$ of the index set $\{1,\ldots, K\}$ denotes the {\em successive encoding order} and where the transmit covariance is given by $\Sigmam_x = \sum_{k=1}^K \vv_k \vv_k^\herm q_k$, defined by the unit-norm ``steering vectors'' $\{\vv_k\}$ and by the users transmit powers $\{q_k\}$.

The transmitter parameters $\{\vv_k\}$, $\{q_k\}$, $\pi$, achieving points on the boundary of
$\Cc(\Sc; \hv_1,\ldots, \hv_K)$, can be determined by
solving the Weighted Rate Sum Maximization (WSRM) problem
\begin{eqnarray} \label{wsrm}
\mbox{maximize} & & \sum_{k=1}^K w_k R_k \nonumber \\
\mbox{subject to} & & \Rm \in \Cc(\Sc; \hv_1,\ldots, \hv_K)
\end{eqnarray}
for some suitable nonnegative weights $\{w_k\}$.
Although a direct solution of (\ref{wsrm}) is difficult, for the special case where the constraint set $\Sc$ is defined by {\em linear inequalities}
\begin{equation} \label{lin-constr}
\trace \left (\Sigmam_x \Phim_\ell \right ) \leq \gamma_\ell,  \;\;\; \ell = 1,\ldots, L,
\end{equation}
where $\{\Phim_\ell\}$ are positive semidefinite symmetric matrices and $\{\gamma_\ell\}$ are non-negative coefficients,
the solution of (\ref{wsrm}) can be computed efficiently by solving a sequence of convex problems.
Explicit algorithms for this computation will be presented in Section \ref{sec:dpc}.

By the Heine-Borel theorem, the compactness of the set $\Sc$ implies that $\Sc$ is {\em bounded} with respect to the Frobenius norm.
Hence, without loss of generality, we can always include an additional trace constraint $\trace(\Sigmam_x) \leq P$ for some sufficiently large $P$,
without modifying the problem. It should also be noticed that (\ref{lin-constr})
includes some particularly important special cases studied in the literature: for $L =1$, $\gamma_1 = P$ and $\Phim_1 = \Id$ we have the
classical sum-power constraint \cite{Caire-Shamai-TIT03, Viswanath-Tse-TIT2003, Vishwanath-Jindal-Goldsmith-TIT03, Yu-Cioffi-TIT04};
for $L = M$ and $\Phim_\ell$ being all zero but one ``1'' in the $(\ell,\ell)$-th position, we have the per-antenna constraint \cite{Yu-Lan-TSP07};
for $L < M$ and $\Phim_\ell$ having all zeros but a segment of  consecutive ``1'' on the diagonal we have the per-group of antennas constraint \cite{Yu-Lan-TSP07};
for some arbitrary $L$ and rank-1 $\Phim_\ell = \cv_\ell \cv_\ell^\herm$ we have a general ``interference'' constraint where the unit-vector $\cv_\ell$ denotes a ``forbidden'' direction along which the transmit power must be not larger than $\gamma_\ell$ \cite{Zhang-Poor-arXiv08}.


{\em Linear beamforming} is a simple precoding strategy that can be an attractive alternative to DPC.
In this case, the achievable rate region has the same form of (\ref{cap-region}) but the encoding order $\pi$ is irrelevant and the sum in the denominator of the term inside the log includes all $i \neq k$. The optimization of the transmit powers $\{q_k\}$, however, is actually more difficult than with DPC since the WSRM problem with linear beamforming has no general convex programming equivalent.
We shall focus on linear Zero-Forcing Beamforming (ZFBF) since in the regime of high SNR it is close to optimal and, as we will see, it lends itself to an efficient solution. In this case, the WSRM problem subject to general linear constraints is given by
\begin{eqnarray} \label{wsrm-zfbf}
{\rm maximize} &  & \sum_{k=1}^K w_k \log\left (1 +  |\hv_k^\herm \vv_k|^2 q_k \right ) \nonumber \\
\mbox{subject to} &  & \hv_j^\herm \vv_k = 0 \;\;\;\; \forall j \neq k \nonumber \\
& & \trace \left (\Sigmam_x \Phim_\ell \right ) \leq \gamma_\ell, \;\;\; \forall \ell
\end{eqnarray}
We assume $K \leq M$ and $\Hm = [\hv_1,\ldots, \hv_K] \in \CC^{M \times K}$ of rank $K$, otherwise the problem is infeasible.
In practical applications, the number of users may be larger than the number of antennas and some greedy user selection algorithm
takes care of selecting an ``active subset'' of size not larger than $M$, but we do not consider this aspect here.
Again, a direct solution of (\ref{wsrm-zfbf}) is difficult. The problem has been addressed using convex
relaxation and the theory of generalized inverses in \cite{Wiesel-Eldar-Shamai-CISS07},
for the case of per-antenna power constraint and equal weights (maximization of the sum-rate).
In Section \ref{sec:zf} we present a new efficient algorithm that addresses (\ref{wsrm-zfbf}) in full generality.

\section{WSRM algorithms for DPC} \label{sec:dpc}

Without loss of generality, assume $w_1 \geq \cdots \geq w_K > 0$.  Then, it is well-known that the optimal DPC encoding order is $\pi = \{1,\ldots,K\}$.
In \cite{Zhang-Poor-arXiv08}, using a technique called ``SINR duality'', the following fundamental results are proved.
Define the ``dual MAC'' corresponding to (\ref{general-down}) as the multiple-access Gaussian channel
\begin{equation} \label{dual-mac}
\yv = \sum_{k=1}^K \hv_k x_k + \zv
\end{equation}
where $\yv, \zv \in \CC^M$, $\zv \sim \Cc\Nc(\zerov, \Sigmam_z(\lambdav))$ with $\Sigmam_z(\lambdav)
= \sum_{\ell = 1}^L \lambda_\ell \Phim_\ell$ for some vector of non-negative coefficients $\lambdav \geq \zerov$ and each
transmitter has power constraint $\EE[|x_k|^2] \leq p_k$, subject to a total sum-power constraint
\begin{equation} \label{dual-mac-constr}
\sum_{k=1}^K p_k \leq \sum_{\ell=1}^L \lambda_\ell \gamma_\ell
\end{equation}
Then, for any $\lambdav \geq \zerov$, the value of the original MIMO BC WSRM problem is upperbounded by the value of the new MAC WSRM problem
\begin{eqnarray} \label{wsrm-dual}
\mbox{maximize} & & \sum_{k=1}^K w_k \widehat{R}_k \nonumber \\
\mbox{subject to} & & \widehat{\Rm} \in \Cc_{\rm MAC}
\end{eqnarray}
where $\Cc_{\rm MAC}$ denotes the capacity region of the dual MAC defined above for given parameters $\lambdav$, $\{\Phim_\ell\}$ and $\{\gamma_\ell\}$. The solution of (\ref{wsrm-dual}) is achieved by {\em successive decoding} in the order $K, K-1, \ldots, 1$, i.e. the reverse of the
DPC encoding order.
Furthermore, the upperbound provided by the dual MAC is tight: denoting by $g(\lambdav)$ the value of the dual-MAC problem for given $\lambdav$,
the value of the MIMO BC problem can be obtained by minimizing $g(\lambdav)$ with respect to $\lambdav \geq 0$.
Hence, the MIMO BC WSRM problem can be solved by iterating between one ``outer problem'' solving the minimization of $g(\lambdav)$
and an ``inner problem'' solving (\ref{wsrm-dual}) for fixed $\lambdav$.
An efficient solution of the inner problem is obtained, with minor modifications, using the Lagrange duality approach of \cite{Yu-TIT06},
as done for example in \cite{Kobayashi-Caire-ICASSP07}.

The outer problem can be solved by a subgradient iteration.
Let $\lambdav(n)$ denote the current value of $\lambdav$ at step $n$.
Then, the next value is given by $\lambdav(n+1) = \lambdav(n) - \epsilon_n \; \sv(\lambdav(n))$,
where $\sv(\lambdav(n))$ is a subgradient of $g(\lambdav)$ at $\lambdav = \lambdav(n)$ and $\epsilon_n = \epsilon_0 \frac{1 + b}{n + b}$ is the adaptation step for some suitable $\epsilon_0, b > 0$.
A subgradient for the problem is given by the vector with components \cite{Zhang-Poor-arXiv08}
$s_\ell(\lambdav) = \gamma_\ell - \trace \left ( \Sigmam_x(\lambdav) \Phim_\ell \right )$, where $\Sigmam_x(\lambdav)$ denotes the transmit covariance
matrix of the MIMO BC corresponding to the dual MAC at given $\lambdav$.
Intuitively, if the $\ell$-th constraint is violated, i.e.,  if  $\gamma_\ell - \trace \left ( \Sigmam_x(\lambdav) \Phim_\ell \right ) < 0$, the corresponding variable  $\lambda_\ell$ must be increased, otherwise, $\lambda_\ell$ is decreased.
The calculation of the subgradient requires the mapping of the solution of the dual MAC (for given $\lambdav$) into
the corresponding solution (powers and steering vectors) of the MIMO BC in order to determine $\Sigmam_x(\lambdav)$.
This is obtained by well-known ``MAC-to-BC'' transformations \cite{Vishwanath-Jindal-Goldsmith-TIT03}.

In \cite{Yu-Lan-TSP07}, the per-antenna power constraint is considered and a ``min-max duality'' approach is used in order to
obtain a saddle-point convex-concave optimization problem that can be solved by an iterative {\em infeasible-start Newton method} \cite{Boyd-Vandenberghe-2004}. Following a similar approach, after some algebra omitted here for lack of space, we find a min-max dual MAC problem for the case of general linear constraints in the form:
\begin{eqnarray} \label{WSRM-interfc-dual-sumP}
& \min_{\lambdav \geq 0} \; \max_{\pv \geq 0}  \;\;\; \sum_{k=1}^K w_k \log \frac{\left | \Sigmam'_z(\lambdav) + \sum_{j=1}^k \hv_j \hv^\herm_j p_j \right |}{\left | \Sigmam'_z(\lambdav) + \sum_{j=1}^{k-1} \hv_j\hv^\herm_j p_j \right |}   & \nonumber \\
& \mbox{subject to} \;\;\; \Sigmam'_z(\lambdav) = \Id + \sum_{\ell=1}^L \lambda_\ell \Phim_\ell, & \nonumber \\
& \sum_k p_k \leq P + \sum_{\ell=1}^L \lambda_\ell \gamma_\ell  &
\end{eqnarray}
where $P$ denotes the total sum-power constraint of the MIMO BC. As we already argued, a sum-power constraint
corresponding to $\Phim_0 = \Id$ and $\gamma_0 = P$ can always be included without loss of generality.
It is not difficult to show that the optimal value of the corresponding dual variable is $\lambda_0 = 1$.
It follows that (\ref{wsrm-dual}) and (\ref{WSRM-interfc-dual-sumP}) are indeed identical.

The infeasible start Newton method can be used as an alternative to the (inner) Lagrange duality -- (outer) subgradient method reviewed before.
Since this algorithm is only briefly presented in \cite{Yu-Lan-TSP07} for the case of per-antenna power constraint,
and several computation steps are left to the reader, we give more details here for the general linear constraint case.
First, we define the modified objective function for (\ref{WSRM-interfc-dual-sumP})
\begin{eqnarray} \label{mac-obj}
f_t(\pv,\lambdav) & = & \sum_{k = 1}^K \Delta_k \log \left | \Id + \sum_{\ell=1}^L \lambda_\ell \Phim_\ell + \sum_{j=1}^k \hv_j \hv_j^\herm p_j \right | - \nonumber \\
& & - w_1 \log \left | \Id + \sum_{\ell=1}^L \lambda_\ell \Phim_\ell \right | \nonumber \\
& & + \frac{1}{t} \left ( \sum_{k=1}^K \log p_k - \sum_{\ell=1}^L \log \lambda_\ell \right )
\end{eqnarray}
where $t > 0$ is a parameter that controls a ``logarithmic barrier'' term in order to prevent the iterative algorithm
to approach the boundaries where some elements in $\pv$ or in $\lambdav$ may become zero or negative and where we define $\Delta_k = w_k - w_{k+1}$ with $w_{K+1} = 0$. The logarithmic barrier guarantees that the optimal value of the problem can be approached with
gap $\frac{K + L}{t}$.  Along the iterations, the value $t$ shall be increased in order to make this gap as small as desired.

The problem is convex with respect to $\lambdav$ and concave with respect to $\pv$, with Lagrangian function  (neglecting the non-negativity constraints and using the modified objective function (\ref{mac-obj})) given by
\begin{equation} \label{lagrange1}
\Lc(\pv,\lambdav, \mu) = f_t(\pv,\lambdav) - \mu \left ( \onev^\transp \pv - P - \gammav^\transp \lambdav \right )
\end{equation}
with $\gammav = (\gamma_1,\ldots,\gamma_L)^\transp$ and $\pv = (p_1,\ldots,p_K)^\transp$.
The necessary and sufficient conditions for optimality are given by the KKT conditions:
\begin{eqnarray} \label{KKT-residual}
\rv_1 & = & \frac{\partial f_t(\pv,\lambdav)}{\partial \pv} - \mu \onev = 0 \nonumber \\
\rv_2 & = & \frac{\partial f_t(\pv,\lambdav)}{\partial \lambdav} + \mu \gammav = 0 \nonumber \\
r_3    & = &  P  + \gammav^\transp \lambdav - \onev^\transp \pv = 0
\end{eqnarray}
The vector $\rv = (\rv_1^\transp, \rv_2^\transp, r_3)^\transp$ of dimension $K + L + 1$ forms the ``residual''
of the KKT equations. The algorithm finds a direction and a step for updating the variables $(\pv, \lambdav, \mu) \geq 0$  such that, as the number of iterations grows,  the norm of the residual tends to zero. The updating direction is given by  $\dv = - \left ( \nabla \rv \right )^{-1} \rv$, where the so-called {\em KKT matrix} is given by
\begin{equation} \label{kktmatrix}
\nabla \rv = \left [ \begin{array}{ccc}
\frac{\partial \rv_1}{\partial \pv^\transp} & \frac{\partial \rv_1}{\partial \lambdav^\transp} & - \onev \vspace{3mm} \\
\frac{\partial \rv_2}{\partial \pv^\transp} & \frac{\partial \rv_2}{\partial \lambdav^\transp} & \gammav \vspace{3mm} \\
- \onev^\transp & \gammav^\transp & 0 \end{array} \right ]
\end{equation}
Letting for  simplicity the vector of variables be denoted by $\xsf = (\pv^\transp, \lambdav^\transp, \mu)^\transp$, the algorithm takes the following form:
\begin{enumerate}
\item Fix the algorithm parameters $\nu > 1$, and $\delta > 0$. Initialize $\xsf(0)$ to some positive values,
let $n = 0$, and $t = 1$.
\item Compute the updating direction $\dv(n)$ at $\xsf(n)$.
(see explicit expressions of the derivatives given later on).
\item Update $\xsf(n+1) = \xsf(n) + s \dv(n)$ where $s$ is found by backtracking line search:
initialize $s = 1$ and find $s$ such as, while
\[ \left \|\rv(\xsf(n) + s \dv(n)) \right \| > (1 - \alpha s) \left \|\rv(\xsf(n)) \right \| \]
then $s \leftarrow \beta s$, where $\beta \in (0,1)$ and $\alpha \in (0,1/2)$ are fixed constants.
(typical values are $\alpha = 0.3$ and $\beta = 0.8$).
\item If $\|\rv(\xsf(n+1)) \| \leq \delta$, move to the next step,
otherwise set $n \leftarrow n + 1$ and go back to step 2.
\item If $\frac{K+L}{t} \leq \delta$, exit and accept the value of $\xsf(n+1)$ as the final value, otherwise
set $t \leftarrow \nu t$ and $n \leftarrow n + 1$ and go back to step 2.
\end{enumerate}
Explicit expressions for the elements of the KKT matrix $\nabla \rv$ can be obtained using matrix calculus
(see for example \cite{Brewer-TCS78} and references therein).
Letting $\Am_k = \left [\Id +  \sum_{\ell=1}^L \lambda_\ell \Phim_\ell + \sum_{j=1}^k \hv_j \hv_j p_j \right ]^{-1}$, we find:\footnote{Here $\delta_{i,j}$ denotes
Kronecker's delta, equal to 1 if $i=j$ and to 0 otherwise.}
\begin{eqnarray*}
\left [ \frac{\partial \rv_1}{\partial \pv^\transp} \right ]_{i,j} & = &  - \sum_{k = \max\{i,j\}}^K \Delta_k  \hv_i^\herm \Am_k \hv_j \hv_j^\herm \Am_k \hv_i  - \frac{\delta_{i,j}}{t p_i^2} \\
\left [ \frac{\partial \rv_1}{\partial \lambdav^\transp} \right ]_{i,j}
& = & - \sum_{k = i}^K \Delta_k \hv_i^\herm \Am_k \Phim_j \Am_k \hv_i = \left [ \frac{\partial \rv_2}{\partial \pv^\transp} \right ]_{j,i} \\
\left [ \frac{\partial \rv_2}{\partial \lambdav^\transp} \right ]_{i,j} & = & - \sum_{k = 1}^K \Delta_k \trace \left ( \Am_k \Phim_j \Am_k \Phim_i \right ) \\
& & + w_1 \trace \left ( \Am_0 \Phim_j \Am_0 \Phim_i  \right ) + \frac{\delta_{i,j}}{t \lambda_i^2}
\end{eqnarray*}

\begin{figure}
  \centering
  \includegraphics[width=3.5in]{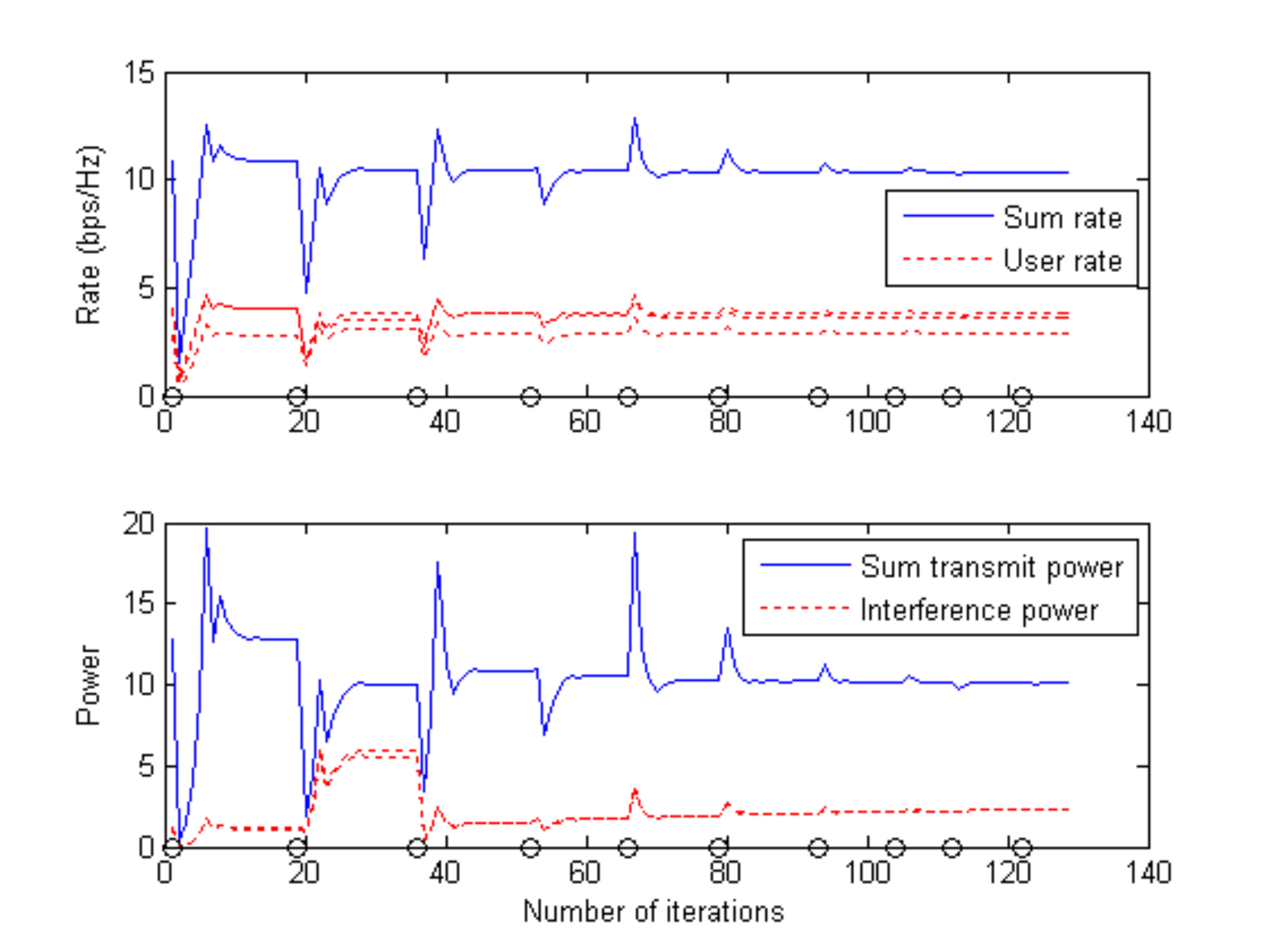}
  \caption{Rate and power convergence behavior of the subgradient method for DPC with $M=4$ and $K=3$ under the sum transmit power and
  interference constraints with $L=2$ forbidden directions.}
  \label{fig:dpcsubgr}
\end{figure}

\begin{figure}
  \centering
  \includegraphics[width=3.5in]{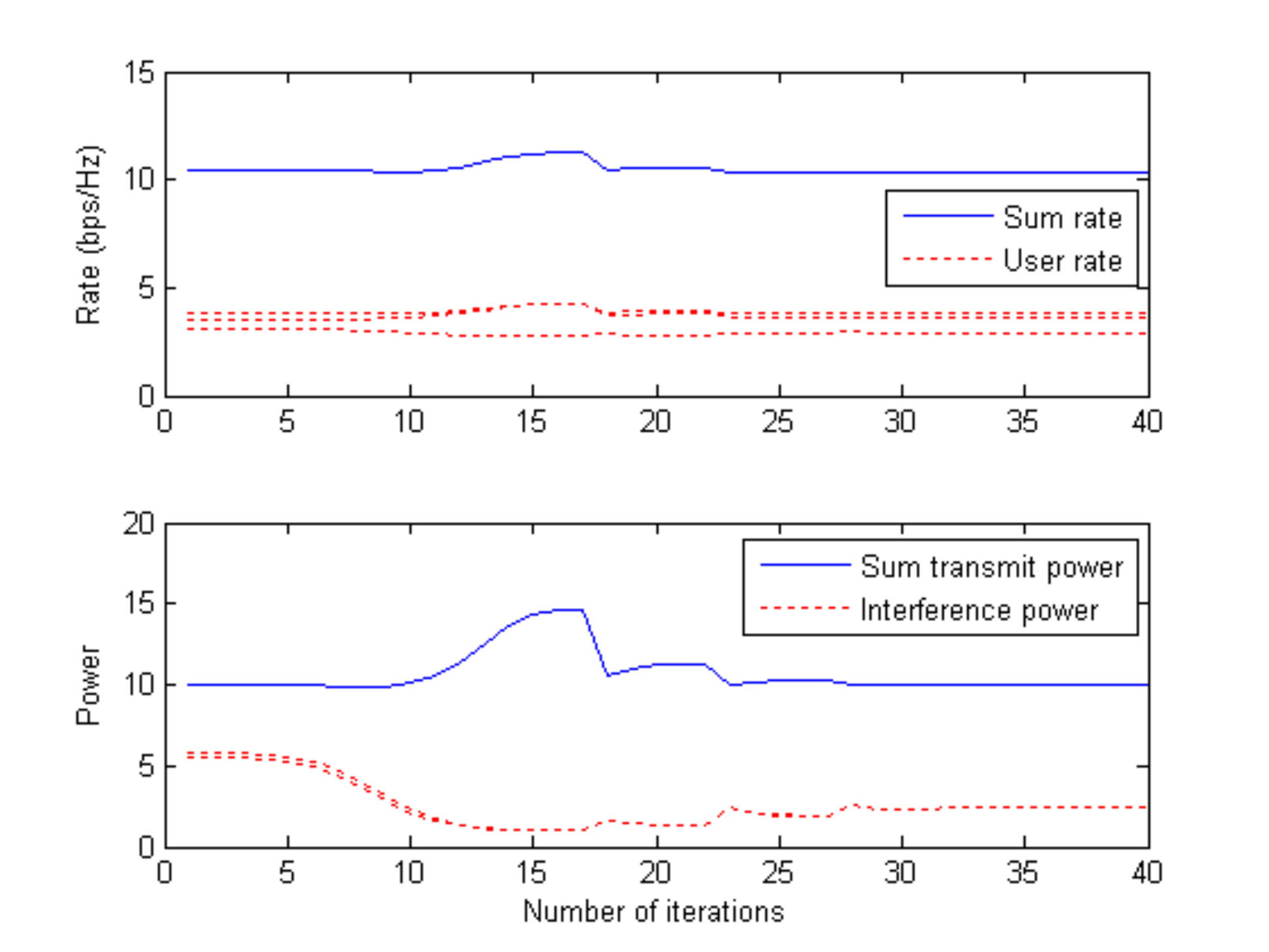}
  \caption{Rate and power convergence behavior of the infeasible start Newton method for DPC under the same conditions of Fig.~\ref{fig:dpcsubgr}.}
  \label{fig:dpcnt}
\end{figure}

Figs.~\ref{fig:dpcsubgr} and \ref{fig:dpcnt} show a numerical example for $M = 4$ antennas, $K = 3$ users, and unit weights for users $w_k = 1$, with $L = 2$ forbidden interference directions.
We considered a sum-power constraint with $P=10$ and interference constraints with $\gamma_1 = \gamma_2 = 2.5$.
The same channel vectors $\hv_k$ and unit-norm interference direction vectors $\cv_\ell$ were used in both plots.
Fig.~\ref{fig:dpcsubgr} and Fig.~\ref{fig:dpcnt} show the evolution of the objective function (sum rate) and of the sum-power and interference constraints versus the number of iterations $n$ for the subgradient-based algorithm and infeasible start Newton algorithm, respectively.
Circles in Fig.~\ref{fig:dpcsubgr} mark the iterations at which the subgradient update is performed, i.e. the start of the new inner problem loop.
Of course, both algorithms converge to the same optimal values and satisfy the given sum-power and interference power constraints. However,
the infeasible start Newton algorithm converges significantly faster.

\section{A novel WSRM algorithm for ZFBF} \label{sec:zf}

The WSRM problem with ZFBF (\ref{wsrm-zfbf}) can be reformulated in terms of unnormalized transmit matrices (i.e., {\em including} the transmit powers) as
\begin{eqnarray} \label{wsrm-zfbf1}
\mbox{maximize} &  & \sum_{k=1}^K w_k \log\left (1 +  \hv_k^\herm \Tm_k \hv_k\right ) \nonumber \\
\mbox{subject to} &  & \hv_j^\herm \Tm_k \hv_j = 0 \;\;\;\; \forall j \neq k \nonumber \\
& & \trace \left (\sum_{k=1}^K \Tm_k \Phim_\ell \right ) \leq \gamma_\ell, \;\;\; \forall \ell  \nonumber \\
&  & \Tm_k \geq 0, \;\;\; {\rm rank}(\Tm_k) = 1, \;\;\; \forall k
\end{eqnarray}

Problem (\ref{wsrm-zfbf}) is not convex due to the rank-1 constraint. In \cite{Wiesel-Eldar-Shamai-CISS07} the problem is solved for the equal-weight case and per-antenna constraint and it is shown that the convex relaxation problem obtained by removing the rank-1 constraints has always a rank-1 solution. Following in the footsteps, it is easy to show that the same holds for the general case (\ref{wsrm-zfbf1}). In particular,  letting $\{\Tm_k^\star\}$ denote a solution of the relaxed problem with possibly rank$(\Tm_k^\star) > 1$ for some $k$, a rank-1 solution $\Tm_k = \tv_k \tv_k^\herm$ achieving the same optimal value can be determined by solving, independently for each user, the problem:
\begin{eqnarray} \label{socp}
\mbox{maximize} && \hv_k^\herm \tv_k \nonumber \\
\mbox{subject to} && \hv_k^\herm \tv_k \in \RR_+ \nonumber \\
&& \hv_j^\herm \tv_k = 0, \;\;\; \forall \; j \neq k \nonumber \\
&& \trace \left ( \tv_k \tv_k ^\herm \Phim_\ell \right ) \leq \trace \left (\Tm_k^\star \Phim_\ell \right ), \;\;\; \forall \; \ell
\end{eqnarray}
We notice that (\ref{socp}) is a Second-Order Cone Program (SOCP) and can be easily solved by standard tools. In the special case of per-antenna constraints, treated in \cite{Wiesel-Eldar-Shamai-CISS07}, (\ref{socp})
reduces to a linear program.

Two main issues arise from the convex relaxation approach. 1) A dramatic dimensionality increase: the relaxed problem deals with $K$ symmetric matrices of dimension $M \times M$, that is,  with $K M(M-1)/2 = O(KM^2)$ variables.
2) Lack of an efficient computational method:
in \cite{Wiesel-Eldar-Shamai-CISS07}, the relaxed problem for equal weights
is cast as a MAXDET for which efficient solvers exist. For general weights, the problem is not MAXDET and general-purpose
convex optimizers must be used, with consequent increase of the computation burden.
In the following we address both issues.

The zero-forcing constraints $\hv_j^\herm \tv_k = 0$ for all $j \neq k$ imply that the linear precoding matrix
$\Tm = [\tv_1, \ldots, \tv_K]$ must be a right generalized inverse of the channel matrix $\Hm^\herm$, i.e., it can be expressed in the form
\begin{eqnarray} \label{T-form}
\Tm & = & [\gv_1 a_1, \ldots, \gv_K a_k] + \Um^\perp [\bv_1, \ldots, \bv_K] \nonumber \\
& = & \Gm + \Um^\perp \Bm
\end{eqnarray}
where $\gv_k$ is the normalized (to unit-norm) $k$-th column of the Moore-Penrose (right) pseudoinverse $\Hm (\Hm^\herm \Hm)^{-1}$ of $\Hm^\herm$, $\{a_k\}$ are scalar coefficients, $\Um^\perp$ is an orthogonal projector onto the orthogonal complement of the span of the channel vectors $\{\hv_k\}$, and $\{\bv_k\}$ are $(M - K)$-dimensional vectors of coefficients. The direct optimization of the coefficients $\{a_k\}$  and $\{\bv_k\}$ can be obtained by iterating two steps: 1) for fixed steering vectors, optimize the power allocation; 2) for fixed {\em relative} powers on the pseudo-inverse directions, maximize a common scaling factor by optimizing the steering vectors.

{\bf Step 1.} Initialize the steering vectors by $\tv_k = \gv_k$, corresponding to $a_k = 1$ and $\bv_k = \zerov$, for all $k$. The ZFBF power allocation problem for fixed (not necessarily unit-norm) steering vectors is
\begin{eqnarray} \label{fixed-steering}
\mbox{maximize} & & \sum_{k=1}^K w_k \log(1 + |\hv_k^\herm \tv_k|^2 q_k) \nonumber\\
\mbox{subject to:} & & \sum_{k=1}^K q_k \tv_k^\herm \Phim_\ell \tv_k \leq \gamma_\ell, \;\;\; \forall \ell \nonumber \\
& & \qv \geq \zerov
\end{eqnarray}
Defining the $L \times K$ matrix  $\Cm$ with $(\ell,k)$ element $[\Cm]_{\ell,k} = \frac{1}{\gamma_\ell} \tv_k^\herm \Phim_\ell \tv_k$, the constraint can be written as $\Cm \qv \leq \onev$. The Lagrangian for (\ref{fixed-steering}) is
\begin{equation} \label{Lagrangian}
\Lc(\qv, \lambdav) = \sum_{k=1}^K w_k \log(1 + |\hv_k^\herm \tv_k|^2 q_k) - \lambdav^\transp \Cm \qv + \lambdav^\transp \onev
\end{equation}
where $\lambdav \geq 0$ is a vector of dual variables. The KKT conditions for $q_k$ yield the waterfilling-like solution \begin{equation} \label{waterfilling-like}
q_k(\lambdav) = \left [ \frac{w_k}{\lambdav^\transp \cv_k} - \frac{1}{|\hv_k^\herm \tv_k|^2} \right ]_+
\end{equation}
where $\cv_k$ is the $k$-th column of $\Cm$. Using this into $\Lc(\qv,\lambdav)$, we can solve the dual problem by minimizing $\Lc(\qv(\lambdav),\lambdav)$ with respect to $\lambdav \geq 0$. It is immediate to check that for any, $\lambdav' \geq 0$,
\begin{eqnarray} \label{subgradient-zf}
\Lc(\qv(\lambdav'),\lambdav') & \geq & \Lc(\qv(\lambdav),\lambdav') \nonumber \\
& = &  \Lc(\qv(\lambdav),\lambdav) + (\onev - \Cm \qv(\lambdav))^\transp ( \lambdav' - \lambdav) \nonumber \\
& &
\end{eqnarray}
Therefore,  $\sv(\lambdav) = (\onev - \Cm \qv(\lambdav))$ is a subgradient for $ \Lc(\qv(\lambdav),\lambdav)$. It follows that the dual problem can be solved by a simple $L$-dimensional subgradient iteration.

{\bf Step 2.} Let $\{q_k\}$ denote the output of Step 1 for fixed steering vectors $\{\tv_k\}$. It follows that, by construction, $a_k = \sqrt{q_k} \gv_k^\herm \tv_k$. In this step we fix $\{a_k\}$ as given above and search for the steering vectors that maximize a common power scaling factor $\eta$. Using (\ref{T-form}) we obtain the optimization problem
\begin{eqnarray} \label{max-mu}
\mbox{maximize} & & \eta \nonumber\\
\mbox{subject to:} & & \frac{\eta^2}{\gamma_\ell} \trace \left ( \Tm \Tm^\herm \Phim_\ell \right ) \leq 1 \;\;\; \forall \; \ell
\end{eqnarray}
with solution readily given by
\[ \eta = \frac{1}{\max_{\ell=1,\ldots,L} \sqrt{ \frac{1}{\gamma_\ell} \trace \left ( \Tm \Tm^\herm \Phim_\ell \right )}} \]
where $\Bm$ is the solution of
\begin{eqnarray} \label{min-t}
\mbox{minimize} & & u \nonumber\\
\mbox{subject to:} & &  \sqrt{ \frac{1}{\gamma_\ell} \trace \left ( \Tm \Tm^\herm \Phim_\ell \right )} \leq u \;\;\; \forall \; \ell
\end{eqnarray}
and where $\Tm$ and $\Bm$ are related by (\ref{T-form}).
It is recognized that (\ref{min-t}) is a SOCP with respect to the variables $u$ and $\Bm$ and can be solved by standard efficient tools \cite{YALMIP}.

The output of Step 2 is a new set of steering vectors in the form $\tv_k = \eta [ \gv_k a_k + \Um^\perp \bv_k ]$. These can be used as new fixed steering vectors for Step 1, and so on. With the proposed initialization, at the first round of Step 1 we obtain the optimal solution based on the pseudo-inverse steering vectors. Then, the algorithm finds a generalized inverse that improves upon the pseudo-inverse already after one iteration.
Although It is known (see \cite{Wiesel-Eldar-Shamai-CISS07}) that the pseudo-inverse ZFBF is optimal under the sum-power constraint,
under general linear constraints it may be very suboptimal.

\begin{figure}
  \centering
  \includegraphics[width=3.5in]{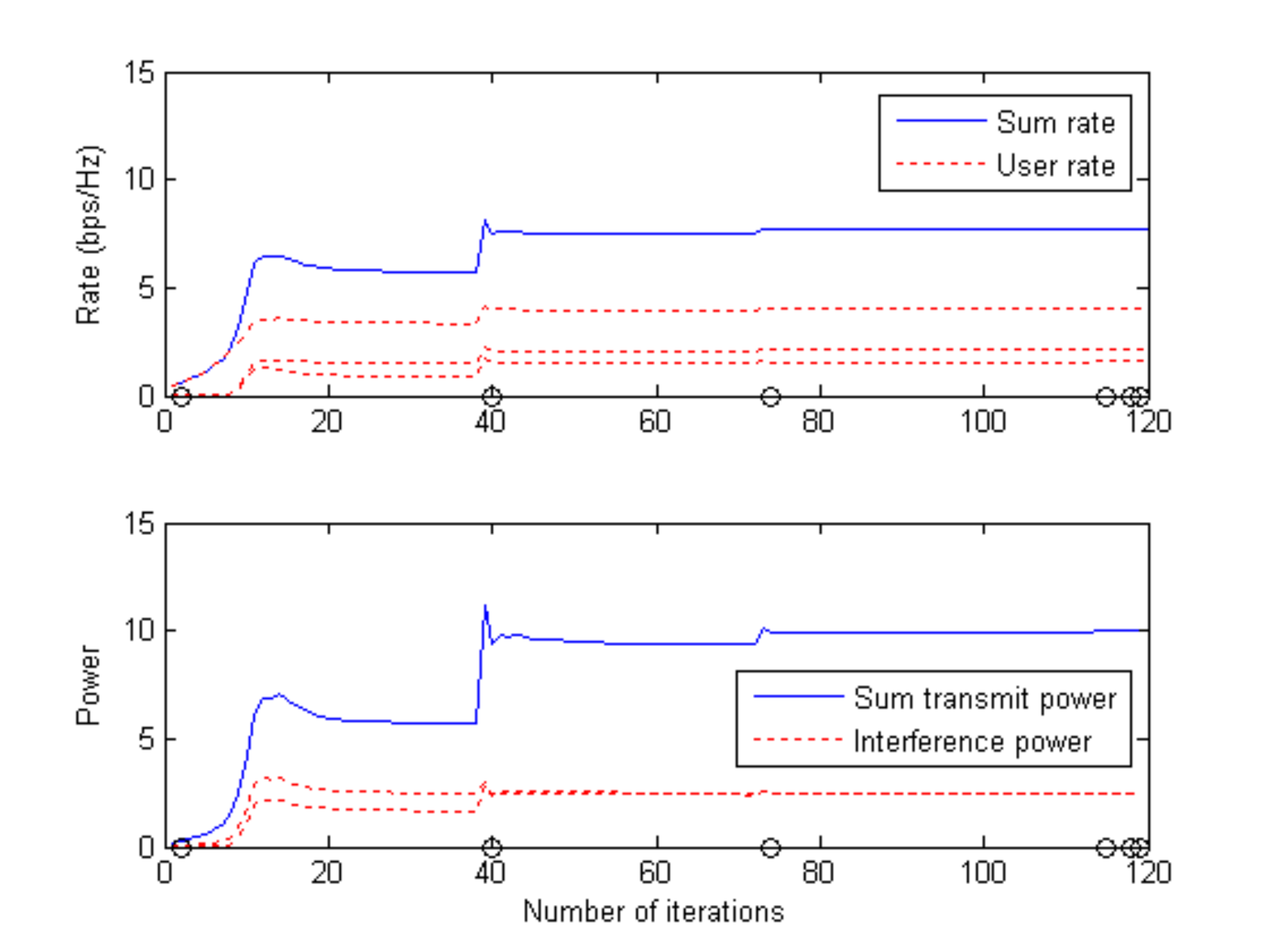}
  \caption{Rate and power convergence behavior for ZF-BF under the same conditions of Fig.~\ref{fig:dpcsubgr}.}
  \label{fig:zfbfgi}
\end{figure}

Fig.~\ref{fig:zfbfgi} illustrates the convergence behavior of the proposed iterative algorithm for ZFBF under general linear constraints.
Channel and constraint parameters are the same as in Figs.~\ref{fig:dpcsubgr} and \ref{fig:dpcnt}.
The proposed algorithm for ZFBF satisfies the given sum-power and interference power constraints as in DPC cases.
The circles indicate the iterations at which the steering vectors update (step 2) is performed, i.e., when the power optimization of step 1 begins with the new
set of steering vectors. Since the steering vectors are initialized with the pseudo-inverse directions, the performance of pseudo-inverse ZFBF
is given at the end of the first round of step 1, i.e. right on the left of the second circle in the plots (iteration number 39 in the plot).
We notice that the pseudo-inverse ZFBF is markedly suboptimal in this case, in fact, the transmit sum power constraint is not met with equality.
This means that if one insists on pseudo-inverse steering vectors the transmitter has to back off its transmit power in order to meet the interference constraints.
Instead, our algorithm finds a generalized inverse that yields a significant improvement and, in this case,
meets all constraints with equality.

\bibliographystyle{IEEEtran}
\bibliography{ISIT2009_CR}

\begin{thebibliography}{10}
\providecommand{\url}[1]{#1}
\csname url@samestyle\endcsname
\providecommand{\newblock}{\relax}
\providecommand{\bibinfo}[2]{#2}
\providecommand{\BIBentrySTDinterwordspacing}{\spaceskip=0pt\relax}
\providecommand{\BIBentryALTinterwordstretchfactor}{4}
\providecommand{\BIBentryALTinterwordspacing}{\spaceskip=\fontdimen2\font plus
\BIBentryALTinterwordstretchfactor\fontdimen3\font minus
  \fontdimen4\font\relax}
\providecommand{\BIBforeignlanguage}[2]{{%
\expandafter\ifx\csname l@#1\endcsname\relax
\typeout{** WARNING: IEEEtran.bst: No hyphenation pattern has been}%
\typeout{** loaded for the language `#1'. Using the pattern for}%
\typeout{** the default language instead.}%
\else
\language=\csname l@#1\endcsname
\fi
#2}}
\providecommand{\BIBdecl}{\relax}
\BIBdecl

\bibitem{Caire-Jindal-Kobayashi-ISIT07}
G.~Caire, N.~Jindal, M.~Kobayashi, and N.~Ravindran, ``{Quantized vs. analog
  feedback for the MIMO broadcast channel: a comparison between zero-forcing
  based achievable rates},'' in \emph{Proc. IEEE Int. Symp. on Inform. Theory,
  ISIT}, Nice, France, June 2007.

\bibitem{Caire-Jindal-Kobayashi-ISIT08}
M.~Kobayashi, G.~Caire, and N.~Jindal, ``{How much training and feedback are
  needed in MIMO broadcast channels?}'' in \emph{Proc. IEEE Int. Symp. on
  Inform. Theory, ISIT}, Toronto, Canada, July 2008.

\bibitem{Weingarten-Steinberg-Shamai-TIT04}
H.~Weingarten, Y.~Steinberg, and S.~Shamai, ``{The capacity region of the
  Gaussian multiple-input multiple-output broadcast channel},'' \emph{IEEE
  Trans. on Inform. Theory}, vol.~52, no.~9, pp. 3936--3964, Sept. 2006.

\bibitem{Caire-Shamai-TIT03}
G.~Caire and S.~Shamai, ``{On the achievable throughput of a multiantenna
  Gaussian broadcast channel},'' \emph{IEEE Trans. on Inform. Theory}, vol.~49,
  no.~7, pp. 1691--1706, 2003.

\bibitem{Viswanath-Tse-TIT2003}
P.~Viswanath and D.~Tse, ``{Sum capacity of the vector Gaussian broadcast
  channel and uplink-downlink duality},'' \emph{IEEE Trans. on Inform. Theory},
  vol.~49, no.~8, pp. 1912--1921, 2003.

\bibitem{Vishwanath-Jindal-Goldsmith-TIT03}
S.~Vishwanath, N.~Jindal, and A.~Goldsmith, ``{Duality, achievable rates, and
  sum-rate capacity of Gaussian MIMO broadcast channels},'' \emph{IEEE Trans.
  on Inform. Theory}, vol.~49, no.~10, pp. 2658--2668, 2003.

\bibitem{Yu-Cioffi-TIT04}
W.~Yu and J.~Cioffi, ``{Sum capacity of Gaussian vector broadcast channels},''
  \emph{IEEE Trans. on Inform. Theory}, vol.~50, no.~9, pp. 1875--1892, 2004.

\bibitem{Yu-Lan-TSP07}
W.~Yu and T.~Lan, ``{Transmitter optimization for the multi-antenna downlink
  with per-antenna power constraints},'' \emph{IEEE Trans. on Sig. Proc.},
  vol.~55, no.~6, pp. 2646--2660, June 2007.

\bibitem{Zhang-Poor-arXiv08}
L.~Zhang, R.~Zhang, Y.~Liang, Y.~Xin, and H.~V. Poor, ``{On Gaussian MIMO
  BC-MAC duality with multiple transmit covariance constraints},'' \emph{posted
  on arXiv:0809.4101/cs.IT}, 2008.

\bibitem{Wiesel-Eldar-Shamai-CISS07}
A.~Wiesel, Y.~Eldar, and S.~Shamai, ``{Optimal generalized inverses for zero
  forcing precoding},'' in \emph{{Proc. Conference on Information Sciences and
  Systems, CISS}}, Baltimore, MD, March 2007.

\bibitem{Yu-TIT06}
W.~Yu, ``{Sum-capacity computation for the Gaussian vector broadcast channel
  via dual decomposition},'' \emph{IEEE Trans. on Inform. Theory}, vol.~52,
  no.~2, pp. 754--759, Feb. 2006.

\bibitem{Kobayashi-Caire-ICASSP07}
M.~Kobayashi and G.~Caire, ``{Iterative water-filling for weighted sum-rate
  maximization in MIMO-OFDM broadcast channels},'' in \emph{Proc. IEEE Int'l
  Conference on Acoustics, Speech, and Signal Processing, ICASSP}, Honolulu,
  HI, April 2007.

\bibitem{Boyd-Vandenberghe-2004}
S.~Boyd and L.~Vandenberghe, \emph{{Convex Optimization}}.\hskip 1em plus 0.5em
  minus 0.4em\relax Cambridge University Press, 2004.

\bibitem{Brewer-TCS78}
J.~W. Brewer, ``{Kronecker products and matrix calculus in system theory},''
  \emph{IEEE Trans. on Circuits and Systems}, vol. CAS-25, no.~9, pp. 772--781,
  Sept. 1978.

\bibitem{YALMIP}
\BIBentryALTinterwordspacing
J.~Lofberg, ``{YALMIP: A toolbox for modeling and optimization in {MATLAB}},''
  in \emph{Proc. of IEEE Int'l Symposium on Computer-Aided Control System
  Design, CACSD}, Taipei, Taiwan, Sept. 2004. [Online]. Available:
  \url{http://control.ee.ethz.ch/~joloef/yalmip.php}
\BIBentrySTDinterwordspacing

\end{thebibliography}

\end{document}